\documentclass[english]{article}
\usepackage[T1]{fontenc}
\usepackage[latin9]{inputenc}
\usepackage{xcolor}
\usepackage{float}
\usepackage{units}
\usepackage{amsmath}
\usepackage{amssymb}
\usepackage{graphicx}
\usepackage{esint}
\PassOptionsToPackage{normalem}{ulem}
\usepackage{ulem}

\makeatletter

\providecolor{lyxadded}{rgb}{0,0,1}
\providecolor{lyxdeleted}{rgb}{1,0,0}

\newcommand{\lyxaddress}[1]{
\par {\raggedright #1
\vspace{1.4em}
\noindent\par}
}

\makeatother

\usepackage{babel}
\begin{document}

\title{Effective noise theory for the Nonlinear Schrödinger Equation with
disorder}

\author{Erez Michaely and Shmuel Fishman}

\maketitle

\lyxaddress{Physics Department, Technion - Israel Institute of Technology, Haifa
32000, Israel}
\begin{abstract}
For the Nonlinear Shrödinger Equation with disorder it was found numerically
that in some regime of the parameters Anderson localization is destroyed
and subdiffusion takes place for a long time interval. It was argued
that the nonlinear term acts as random noise. In the present work
the properties of this effective noise are studied numerically. Some
assumptions made in earlier work were verified, and fine details were
obtained. The dependence of various quantities on the localization
length of the linear problem were computed. A scenario for the possible
breakdown of the theory for a very long time is outlined.
\end{abstract}

\section{Introduction}

The Nonlinear Schrödinger Equation (NLSE) \cite{Sulem1999} in a random
potential takes the form of 
\begin{equation}
i\partial_{t}\psi=H_{0}\psi+\beta\left\vert \psi\right\vert ^{2}\psi,\label{nls}
\end{equation}
 where $H_{0}$ is the linear part with a disordered potential, which
on a lattice takes the form of
\begin{equation}
H_{0}\psi(x)=-\left(\psi(x+1)+\psi(x-1)\right)+\varepsilon(x)\psi(x).\label{eq:linear hamiltonian}
\end{equation}
 In this work it is assumed that $\varepsilon\left(x\right)$ are
identical independent random variables (i.i.d) uniformly distributed
in the interval of $\left[\frac{-W}{2},\frac{W}{2}\right].$ 

The NLSE was derived for a variety of physical systems under some
approximations. It was derived in classical optics where $\psi$ is
the electric field by expanding the index of refraction in powers
of the electric field keeping only the leading nonlinear term \cite{Agrawal2007}.
For Bose-Einstein Condensates (BEC), the NLSE is a mean field approximation
where the term proportional to the density $\beta|\psi|^{2}$ approximates
the interaction between the atoms. In this field the NLSE is known
as the Gross-Pitaevskii Equation (GPE) \cite{Dalfovo1999,Pitaevskii2003,Leggett2001,Pitaevskii1961,Gross1961}.
It is well known that in 1D in the presence of a random potential
with probability one all the states are exponentially localized \cite{Ishii1973,Lee1985,Lifshits1988}.
Consequently, diffusion is suppressed and in particular a wavepacket
that is initially localized will not spread to infinity. This is the
phenomenon of Anderson localization \cite{Anderson1958}. The problem
defined by \eqref{nls} is relevant for experiments in nonlinear optics,
for example disordered photonic lattices \cite{Schwartz2007,Lahini2008},
where Anderson localization was found in presence of nonlinear effects
as well as experiments on BECs in disordered optical lattices \cite{Gimperlein2005,Lye2005,Clement2005,Clement2006,Sanchez-Palencia2007,Billy2008,Fort2005,Akkermans2008,Paul2007,Beilin2010}.
The interplay between disorder and nonlinear effects leads to new
interesting physics \cite{Fort2005,Akkermans2008,Bishop1995,Rasmussen1999,Kopidakis1999,Kopidakis2000}.
In spite of the extensive research, many fundamental problems are
still open (see recent review to be published in Nonlinearity \cite{Fishman2011}).
In particular there is disagreement between the analytical and the
numerical results \cite{Wang2008,Wang2008a,Fishman2009a,Krivolapov2010,Pikovsky2011,Ivanchenko2011,Benettin1988,Johansson2010,Basko2011}. 

A natural question is whether a wave packet that is initially localized
in space will indefinitely spread for dynamics controlled by (\ref{nls}).
A simple argument indicates that spreading will be suppressed by randomness.
If unlimited spreading takes place the amplitude of the wave function
will decay since the $l^{2}$ norm is conserved. Consequently, the
nonlinear term will become negligible and Anderson localization will
take place as a result of the randomness as conjectured by Fröhlich
\emph{et al} \cite{Frohlich1986}. Contrary to this intuition, based
on the smallness of the nonlinear term resulting from the spread of
the wave function, it is claimed that for the kicked-rotor a nonlinear
term leads to delocalization if it is strong enough \cite{Shepelyansky1993}.
It is also argued that the same mechanism results in delocalization
for the model \eqref{nls} with sufficiently large $\beta$, while,
for weak nonlinearity, localization takes place \cite{Shepelyansky1993,Pikovsky2008}.
Recently, it was rigorously shown that the initial wavepacket cannot
spread so that its amplitude vanishes at infinite time, for large
enough $\beta$ \cite{Kopidakis2008}. It does not contradict spreading
of a fraction of the wavefunction. Indeed, subdiffusion was found
in numerical experiments \cite{Shepelyansky1993,Pikovsky2008,Molina1998,Flach2009,Skokos2009}.
It was also argued that nonlinearity may enhance discrete breathers
\cite{Kopidakis1999,Kopidakis2000}. In conclusion, it is \emph{not}
clear what is the long time behavior of a wave packet that is initially
localized, if both nonlinearity and disorder are present \cite{Fishman2011}.
The major difficulty in numerical resolution of this question is integration
of \eqref{nls} to large time. Most researchers who run numerical
simulations use a split-step method for integration, however it is
impossible to achieve convergence for large times, and therefore some
heuristic arguments assuming that the numerical errors do not affect
the results qualitatively, are utilized \cite{Shepelyansky1993,Flach2009}.
Moreover the problem is chaotic, therefore the trajectories that are
found are not the actual trajectories and it is argued that it does
not affect the statistical results.

Recent rigorous arguments \cite{Wang2008,Wang2008a} in the limit
of strong disorder combined with perturbation theory \cite{Fishman2009a,Krivolapov2010,Fishman2008a}
indicate that it is unlikely that sub-diffusion persists forever and
the asymptotic growth is at most logarithmic in time. Also other recent
work based on a scaling theory \cite{Pikovsky2011} and phase space
considerations \cite{Johansson2010,Basko} lead to similar indications.
It is clear that there is a substantial regime in time and parameters
where sub-diffusion may hold and the purpose of the present work is
to analyze the dynamics in this regime. 

Our analysis based on \cite{Flach2009,Skokos2009}, is conveniently
expressed expanding the wavefunction 
\begin{equation}
\psi(x,t)=\sum_{n}c_{n}(t)u_{n}(x)e^{-iE_{n}t}\label{eq:expansion of psi}
\end{equation}
 where $u_{n}$ are the eigenfunctions of $H_{0}$ typically falling
off exponentially: 
\begin{equation}
u_{n}(x)\approx\frac{e^{-|x_{n}-x|/\xi}}{\sqrt{\xi}}\varphi(x)\label{eq:scale of u}
\end{equation}
 where $\varphi(x)$ is a random function of order unity. The localization
center is $x_{n}$. The $c_{n}(t)$ satisfy

\begin{equation}
i\partial_{t}c_{n}(t)=\beta\sum_{m_{1},m_{2},m_{3}}V_{n}^{m_{1},m_{2},m_{3}}e^{i(E_{n}+E_{m_{1}}-E_{m_{2}}-E_{m_{3}})t}c_{m_{1}}^{*}c_{m_{2}}c_{m_{3}}\equiv F_{n}\left(t\right)\label{Cs}
\end{equation}
 and 
\begin{equation}
V_{n}^{m_{1},m_{2},m_{3}}=\sum_{x}u_{n}(x)u_{m_{1}}(x)u_{m_{2}}(x)u_{m_{3}}(x).\label{Vnmmm}
\end{equation}
\\
In \cite{Flach2009,Skokos2009} it is argued that $F_{n}\left(t\right)$
behaves as random noise with rapidly decaying correlation functions.
The implications are analyzed in Sec. 2 and tested numerically in
Sec. 3. A scenario for the breakdown of the effective noise theory
is outlined in Sec. 4. The results are summarized and open question
are presented in Sec. 5.

\section{The effective noise theory\label{sec:The-effective-noise}}

The theory of SKFF (Skokos, Krimer, Komineas and Flach \cite{Flach2009,Skokos2009})
assumes for spreading to take place to the region where the $n$-th
state is localized from the region where the states $m_{1},m_{2},m_{3}$
have a large amplitude: 
\begin{equation}
|c_{m_{1}}|^{2}\approx|c_{m_{2}}|^{2}\approx|c_{m_{3}}|^{2}\approx\rho\label{cm}
\end{equation}
 while 
\begin{equation}
|c_{n}|^{2}\ll\rho\label{cn}
\end{equation}
It is assumed that the RHS of (\ref{Cs}) is a random function denoted
by $F_{n}\left(t\right)$. We turn to estimate its typical behavior.
First we note that the overlap sums (\ref{Vnmmm}) are random functions.
Within the scaling theory for localization one expects that for sufficiently
weak disorder their various moments are determined by the localization
length. For the case where all indices $\left(n,m_{1},m_{2},m_{3}\right)$
are identical the average is just the inverse participation ratio
what is proportional to $\nicefrac{1}{\xi}$. For the general case
the scaling theory suggests it is a function only of $\xi$. Experience
with scaling theories leads us to assume it is a power of $\xi$.
Therefore we try the form, 
\begin{equation}
\left\langle V_{n}^{m_{1},m_{2},m_{3}}\right\rangle =C_{0}^{\left(1\right)}\xi^{-\eta_{1}},\label{eq:eta1}
\end{equation}
and for the second moment we try to fit to, 
\begin{equation}
<|V_{n}^{m_{1},m_{2},m_{3}}|^{2}>=C_{0}^{\left(2\right)}\xi^{-2\eta_{2}}.\label{V}
\end{equation}
Here $C_{0}^{\left(1\right)}$ and $C_{0}^{\left(2\right)}$ are constants
and $<..>$ is an average over realizations. We note that when the
$m_{i}$ and $n$ are all different the average of the overlap integrals
vanishes. We should note that the localization length $\xi$ is actually
energy dependent. For weak disorder in the center of the band, $\xi\sim W^{-2}$
\cite{DerridaB.1984,MacKinnon1981}, this relation holds for most
energies in the energy band \cite{DerridaB.1984}. In what follows
we will estimate the values of $\eta_{1}$ and $\eta_{2}$ for various
disorder strengths and for various sites $(x_{n},x_{m_{1}},x_{m_{2}},x_{m_{3}})$,
which are within the localization length. Otherwise the sum (\ref{Vnmmm})
is negligible. It is not obvious that both (\ref{eq:eta1}) and (\ref{V})
will scale in this way although it is expected from the scaling theory
of localization, that this is the case for sufficiently weak disorder,
namely large $\xi$. We demonstrare that this is indeed the case and
there is a typical magnitude of the value of the of the overlap sum
(\ref{Vnmmm}) and it scales as, 
\begin{equation}
V=C_{1}\xi^{-\eta}\label{Vt}
\end{equation}
 where $C_{1}$ is a constant. Here and in what follows we denote
by $\xi$ the localization length in the center of the band.

Making the assumption that $F_{n}$ is random combined with (\ref{cm})
, the sum on the RHS of (\ref{Cs}) consists of the order of $\xi^{3}$
terms, at least for weak disorder. These are rapidly oscillating in
time, and it is a nonlinear function of the $c_{m_{i}}\left(t\right)$.
Therefore it is suggestive that it can be considered random. This
assumption will be tested in detail in subsection \ref{sub:3.1}.
The RHS of (\ref{Cs}) is assumed to take the form \cite{Skokos2009}
\begin{equation}
F_{n}=V\mathcal{P}\beta\rho^{3/2}f_{n}(t)=\frac{C_{1}}{\xi^{\eta}}\mathcal{P}\beta\rho^{3/2}f_{n}(t)\label{F}
\end{equation}
 where $C_{1}$ is a constant and 
\begin{equation}
\mathcal{P}=A_{0}\beta^{\gamma}\xi^{\alpha}\rho\label{P}
\end{equation}
 is proportional to the number of \textquotedbl{}resonant modes\textquotedbl{},
namely ones that strongly affect the dynamics of the state $n$. Although
it is reasonable to assume that the number of resonant modes is proportional
to the density $\rho$ a strong argument for it is missing, nevertheless
it is consistent with all numerical results \cite{Skokos2009,Flach2009}.
We assume here the form (\ref{P}) where $A_{0}$ is a constant independent
of $\beta$ and $\xi$. In the end of this section we argue that within
these assumption $\gamma=1$ in agreement with the assumption of \cite{Skokos2009,Flach2009}.
The value of $\alpha$ is estimated numerically (see subsection \ref{sub:3.3}).
Under these assumptions \eqref{Cs} reduces to: 
\begin{equation}
i\partial_{t}c_{n}(t)=F_{n}(t)
\end{equation}
 Assuming $F_{n}(t)$ can be considered random with rapidly decaying
correlations, in particular we assume that the distribution function
of $f_{n}\left(t\right)$ is stationary and the integral of correlation
function $C\left(t'\right)=\left\langle f\left(0\right)f\left(t'\right)\right\rangle $,
where $\left\langle ..\right\rangle $ is the average over the random
potential, converges. Integration results in 
\begin{equation}
c_{n}(t)=-i\frac{C_{1}}{\xi^{\eta}}\mathcal{P}\beta\rho^{3/2}\int_{0}^{t}dt'f_{n}(t')
\end{equation}
 Integrating over a time interval which is sufficiently large yields:
\begin{equation}
<|c_{n}(t)|^{2}>=\frac{A_{1}}{\xi^{2\eta}}\mathcal{P}^{2}\beta^{2}\rho^{3}t=A_{1}A_{0}^{2}\beta^{2(\gamma+1)}\rho^{5}\xi^{2\alpha-2\eta}t
\end{equation}
 where $A_{1}$ is a constant. The value of $<|c_{n}(t)|^{2}>$ increases
with time and equilibrium is achieved when it takes the value $\rho$.
Transitions between states of the type of $n$ (states with small
amplitude) are ignored in this model. The required time for equilibration
is 
\begin{equation}
T=\frac{1}{B\xi^{-2}\rho^{4}}\label{T}
\end{equation}
 where we define 
\begin{equation}
B=A_{1}A_{0}^{2}\beta^{2(1+\gamma)}\xi^{2\alpha-2\eta+2}\label{eq:18}
\end{equation}
The equilibration time $T$ varies slowly compared to $t$ (see discussion
after \eqref{eq:Capital T}). In other words there is a separation
of time scales. On the time scale $T$ the system seems to reach equilibrium
by a diffusion process and the density becomes constant in a region
that includes the site $n$. Hence on this time scale it seems to
equilibrate. On a longer time scales, there is an even longer equilibration
time scale, and the resulting diffusion is even weaker. The consistency
of the argument results of the fact that $\frac{dT}{dt}\rightarrow0$
for $t\rightarrow\infty$. Therefore it is assumed that the variations
of $\rho$ and $T$ are slow on the scale of $t$ . This assumption
is checked in the end of this section. The resulting diffusion coefficient
is 
\begin{equation}
D=C\frac{\xi^{2}}{T}=CB\rho^{4}\label{eq:diffusion coefficient}
\end{equation}
where $C$ is a constant. The assumption is that the nonlinear term
generates a random walk with the characteristic steps $T$ and $\xi$
in time and space. At time scales $t\gg T$, there is diffusion and
\begin{equation}
M_{2}=Dt,
\end{equation}
where $M_{1}=\sum x\left|\psi\left(x,t\right)\right|^{2}$ and the
variance $M_{2}=\sum\left(x-M_{1}\right)^{2}\left|\psi\left(x,t\right)\right|^{2}$
are the first and second moments. Since the second moment $M_{2}$
is inversely proportional to $\rho^{2}$ one finds 
\begin{equation}
\frac{1}{\rho^{2}}=A_{2}CB\rho^{4}t\label{eq:ro_D_t}
\end{equation}
 where $A_{2}$ is a constant. Therefore 
\begin{equation}
\frac{1}{\rho^{2}}=\left(A_{2}CBt\right)^{1/3}.\label{eq:1 over ro_sq}
\end{equation}
 The second moment satisfies:

\begin{equation}
M_{2}=\frac{1}{A_{2}^{2/3}}(CBt)^{1/3}\label{eq:m2-1}
\end{equation}
 and

\begin{equation}
T=\frac{1}{B\xi^{-2}\rho^{4}}=\frac{C^{2/3}A_{2}^{2/3}\xi^{2}t^{2/3}}{B^{1/3}}=\frac{C\xi^{2}}{M_{2}}t.\label{eq:Capital T}
\end{equation}
 The density $\rho$ and the equilibration time $T$ change with time
as $\rho\sim t^{-\frac{1}{3}}$ and $T\sim t^{\frac{2}{3}}$. Therefore
for $\frac{d\rho}{dt}\sim t^{-\frac{4}{3}}$ and $\frac{dT}{dt}\sim t^{-\frac{1}{3}}$.
First note that in the long time limit $t\rightarrow\infty$ both
derivatives vanish and $\frac{d\rho}{dt}\ll\frac{dT}{dt}$. Therefore
for the derivation of the equilibration time $\rho$ can considered
constant and on long scales of spreading $T$ and $D$ can be considered
constant. Therefore the theory is consistent for large $t$. Since
in the $NLSE$ $\beta$ appears only via the combination $\beta\left|\psi\left(x\right)\right|^{2}$,
it can appear in \eqref{eq:18} and \eqref{eq:diffusion coefficient}
only in the power $4$ (that is in the combination $\beta^{4}\rho^{4}$)
therefore $\gamma=1$.\\
In the next section this theory will be tested numerically.

\section{Numerical tests for the effective noise theory\label{sec:Numerical-tests-for}}

In this section the theory presented in section 2 is tested numerically.
In subsection \ref{sub:3.1} the distribution of the $F_{n}\left(t\right)$
is computed, in subsection \ref{sub:3.2} the first moments of the
overlap sums are calculated while in subsection \ref{sub:3.3} the
dependence of the second moment $M_{2}$ of \eqref{eq:m2-1} on $\xi$
is evaluated.

\subsection{\label{sub:3.1}Statistical properties of $F_{n}\left(t\right)$}

\begin{figure}[H]
\includegraphics[scale=0.5]{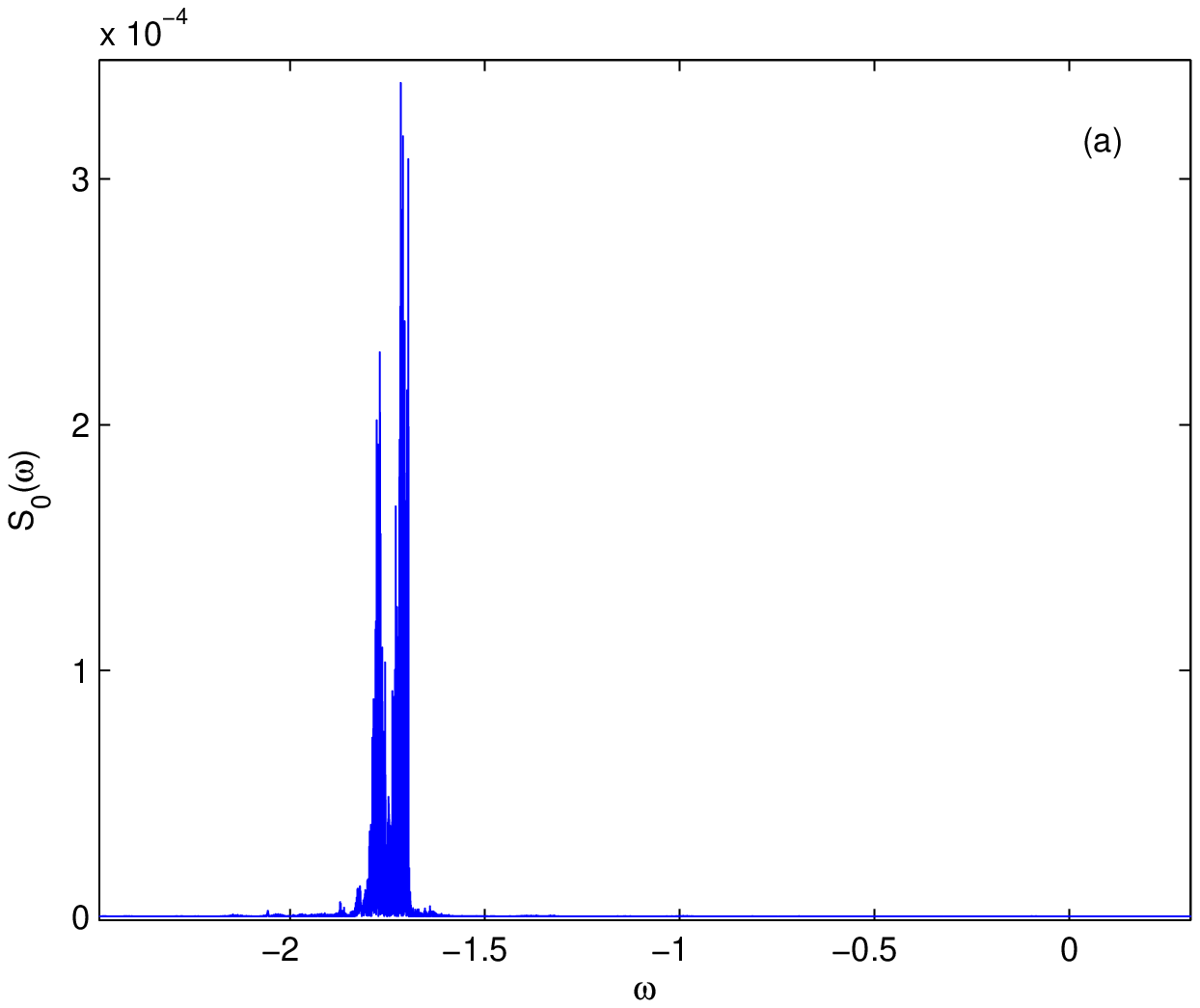}\includegraphics[scale=0.5]{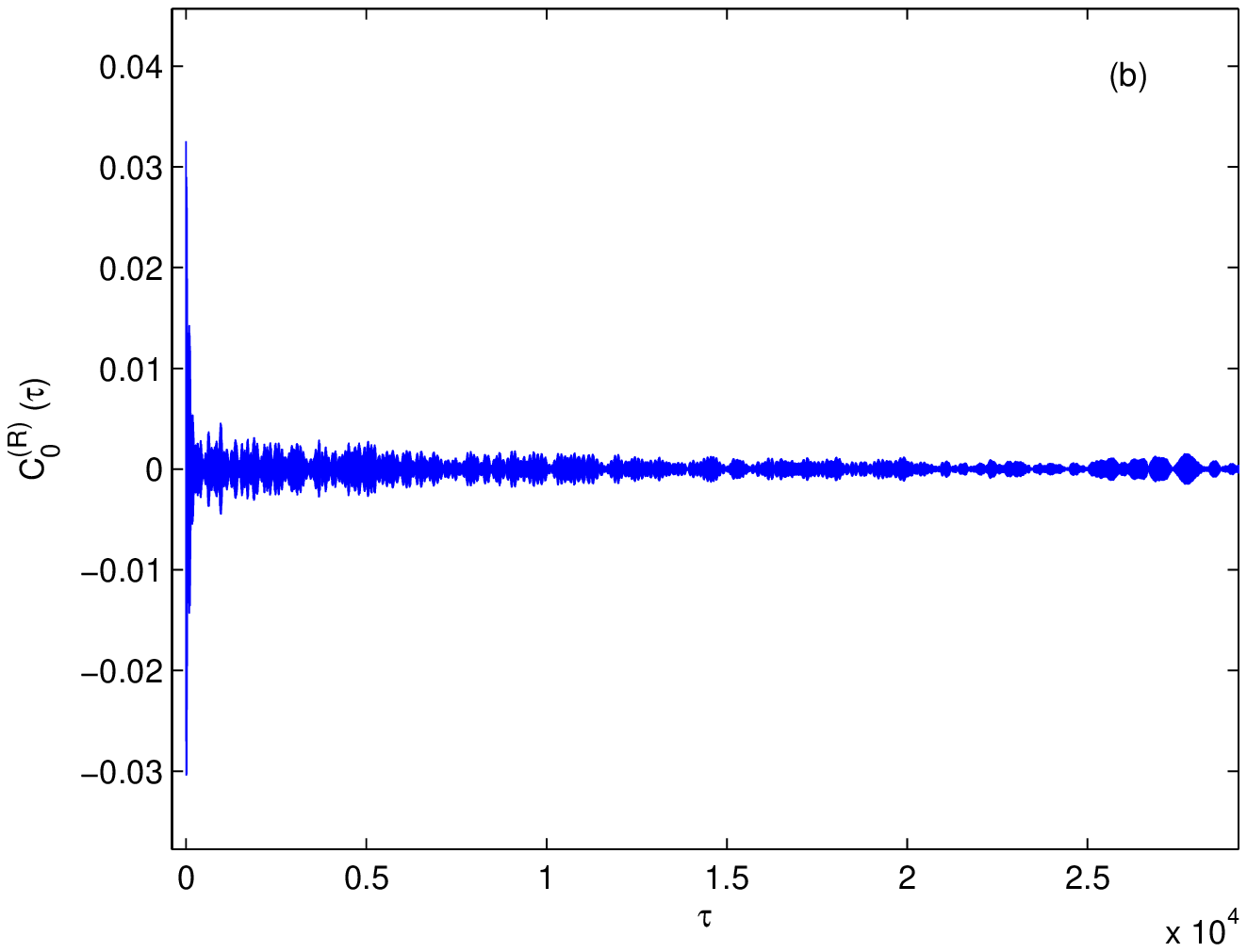}\\
\includegraphics[scale=0.5]{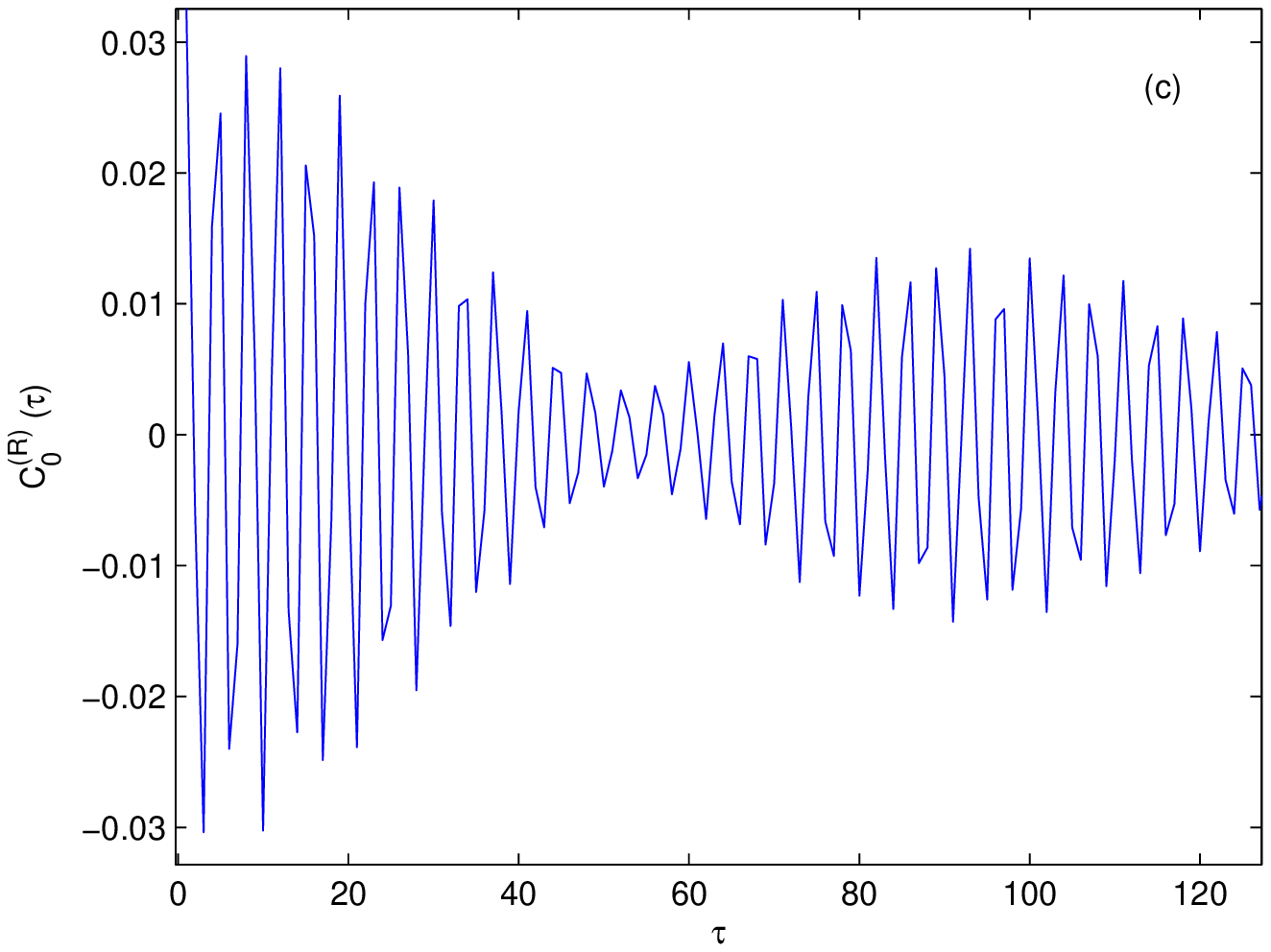}\includegraphics[scale=0.5]{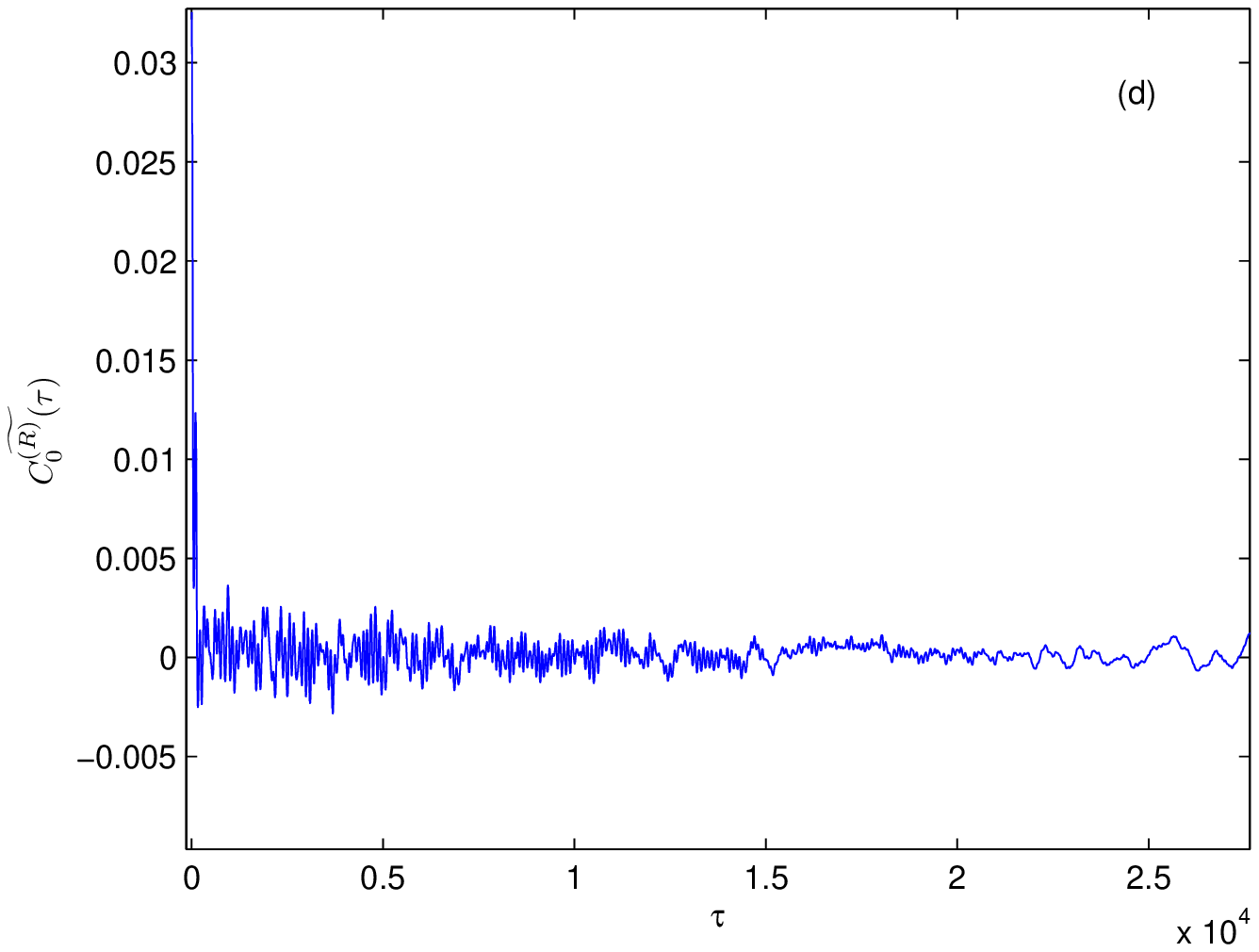}\\
\includegraphics[scale=0.5]{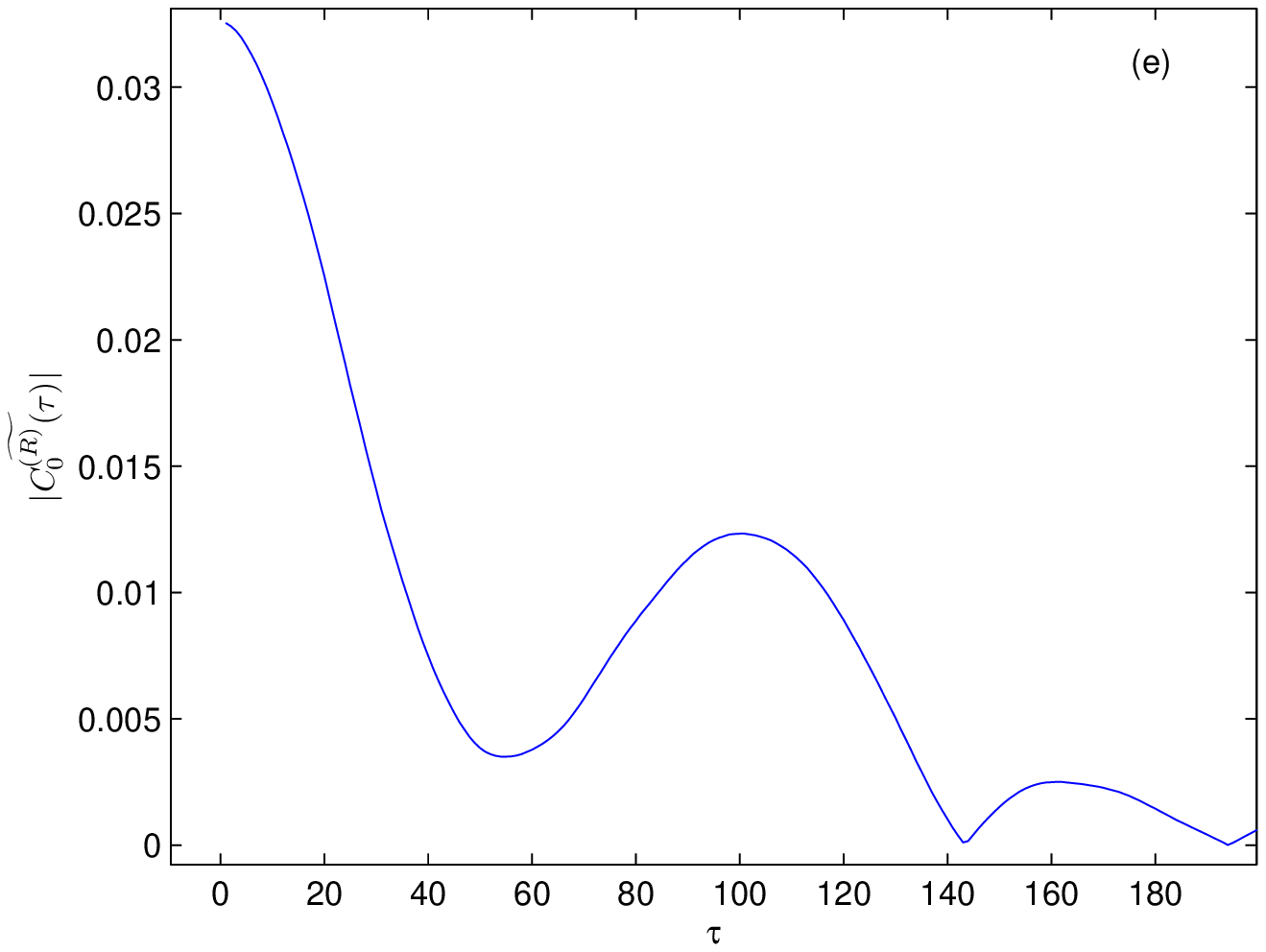}\includegraphics[scale=0.5]{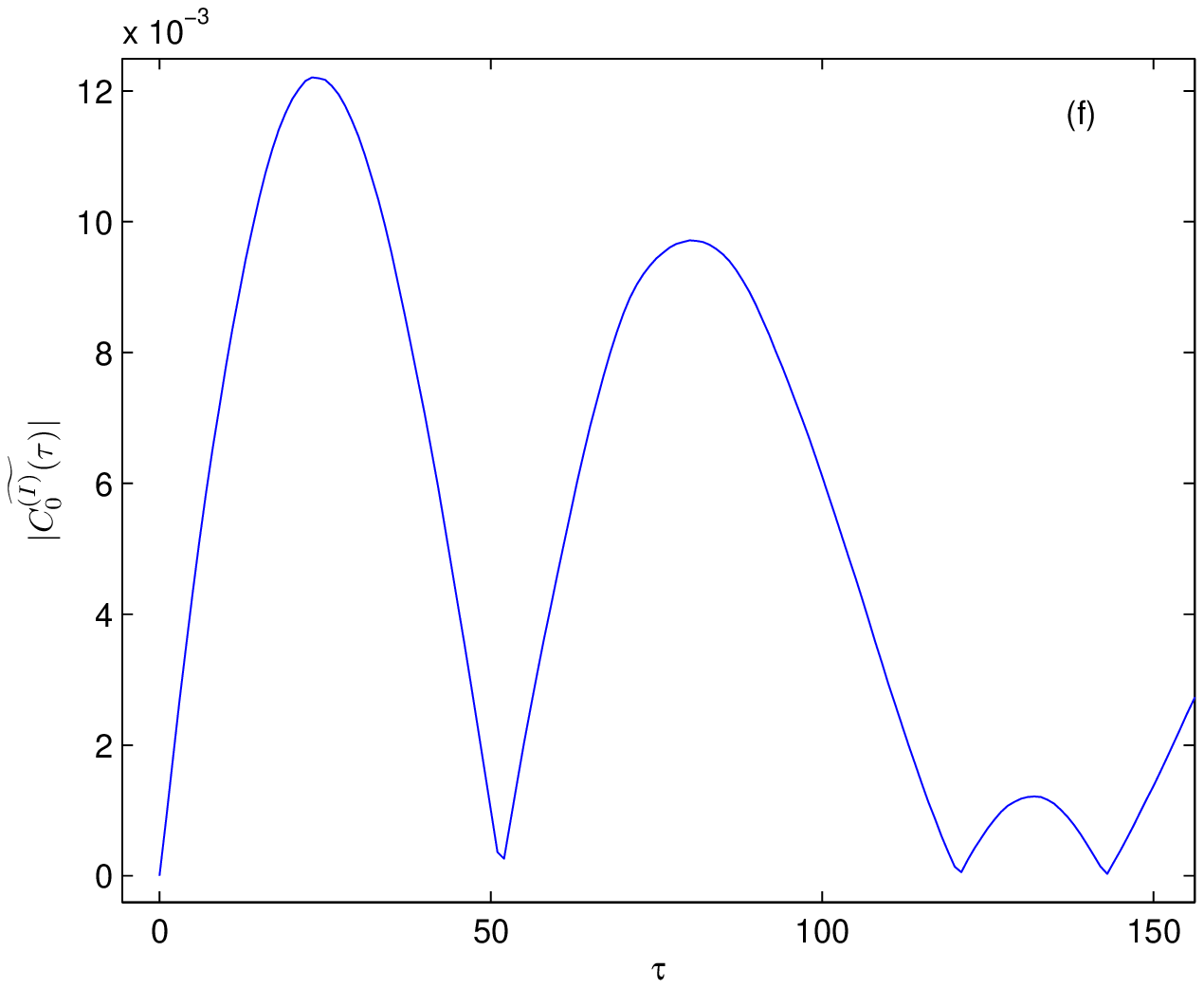}\label{Flo:Distribution W=00003D4 beta=00003D0.1}

\caption{\label{fig:Power Spectrum + A_C_function}The corelation $C_{n}\left(t\right)$
and power spectrum $S_{n}\left(\omega\right)$ of $F_{n}\left(t\right)$
for $W=4$ , $\beta=1$, $N=1024$, $t=10^{5},$$n=0$. (a) The Power
Spectrum $S_{0}\left(\omega\right)$, (b) The auto-correlation function
$C_{0}^{\mbox{\ensuremath{\left(R\right)}}}\left(\tau\right)$, (c)
The zoomed $C_{0}^{\mbox{\ensuremath{\left(R\right)}}}\left(\tau\right)$,
(d) The auto-correlation function $\tilde{C_{0}^{\left(R\right)}}\left(\tau\right)$,
(e) the zoomed $\tilde{C_{0}^{\left(R\right)}}\left(\tau\right)$,
(f) the zoomed $\tilde{C_{0}^{\left(I\right)}}\left(\tau\right)${[}see
text{]}.}
\end{figure}

In this subsection the statistical distribution of $F_{n}\left(t\right)$
is explored. For this purpose the time dependent $NLSE$ (\ref{nls})
was solved numerically for a finite lattice of $N$ sites, for $N_{R}$
realizations of the random potential $\varepsilon\left(x\right)$
and for $W=4$. The wavefunction $\psi\left(x,t\right)$ at time $t$
was calculated for a single site excitation namely the initial condition
$\psi\left(x,0\right)=\delta_{x,0}$ using the split step method.
The details of the numerical calculation are presented in the appendix.
The expansion \eqref{eq:expansion of psi} of $\psi$ in terms of
eigenfunctions of the linear problem \eqref{eq:linear hamiltonian}
yields, 
\begin{equation}
i\partial_{t}c_{n}(t)=\sum_{x}\beta\left|\psi\left(x,t\right)\right|^{2}\psi\left(x,t\right)u_{n}\left(x\right)e^{itE_{n}}\equiv F_{n}\left(t\right).
\end{equation}
 This equation was used to calculate $F_{n}\left(t\right)$ numerically
for a lattice of $N$ sites. In order to check whether $F_{n}\left(t\right)$
can be considered as noise we calculated its power spectrum and auto-correlation
function. First we present results obtained for times up to $t=10^{5}$
for $\beta=1$, $W=4$ ($\xi\approx6.4)$, $N=1024$ for a single
site excitation at $t=0$. The calculation was preformed for $N_{R}=50$
realizations. For nearly all these realizations it was found that
the second moment grows as $M_{2}\propto t^{1/3}$ in agreement with
the results of \cite{Pikovsky2008,Flach2009,Skokos2009}. We focus
first on such realizations and present the results for a specific
realization in Fig. \ref{fig:Power Spectrum + A_C_function} 

The power spectrum is 
\begin{equation}
S_{n}\left(\omega\right)=\left|\hat{F}_{n}\left(\omega\right)\right|^{2},\label{eq:power spectrum}
\end{equation}
where 
\begin{equation}
\hat{F}_{n}\left(\omega\right)=\lim_{\tilde{t}\rightarrow\infty}\frac{1}{\sqrt{\tilde{t}}}\intop_{0}^{\tilde{t}}F_{n}\left(t\right)\cdot e^{\mbox{\ensuremath{\left(-i\omega t\right)}}}dt.\label{eq:Fourier Transform definition}
\end{equation}
It is plotted for some realization in Fig. \ref{fig:Power Spectrum + A_C_function}.a
for n=0. It exhibits a peak around $\left|\omega_{0}\right|\approx1.72$
and its width is $\triangle\omega\approx0.1$. The finite width is
characteristic of noise. Also the Fourier transform of 
\begin{equation}
\tilde{F}_{n}\left(t\right)=F_{n}\left(t\right)\cdot e^{-i\omega_{0}t}\label{eq:deffinition of F_tilde}
\end{equation}
 will exhibit a wide power spectrum near $\omega=0$, with the width
of $\triangle\omega$ that is characteristic of noise. The auto-correlation
function of $F_{n}\left(t\right)$ is 
\begin{equation}
C_{n}\left(\tau\right)=\overline{F_{n}\left(t\right)\cdot F_{n}^{*}\left(t+\tau\right)}\label{eq:a_c_f_definition}
\end{equation}
where bar denotes time average $\overline{g\left(t\right)}\equiv\lim_{\tilde{t}\rightarrow\infty}\frac{1}{\widetilde{t}}\int_{0}^{\widetilde{t}}g\left(t\right)dt$. 

For $\tilde{F}_{n}\left(t\right)$ we define the auto-correlation
function $\tilde{C_{n}}\left(\tau\right)$ that is just \eqref{eq:a_c_f_definition}
with $F_{n}\left(t\right)$ replaced by $\tilde{F}_{n}\left(t\right)$
. In Fig. \ref{fig:Power Spectrum + A_C_function} .b we plot $C_{n}^{\left(R\right)}=Re\left(C_{n}\left(\tau\right)\right)$
for $n=0$ while in Fig.\ref{fig:Power Spectrum + A_C_function} c
the zoomed version is plotted. Note an oscillation of frequency of
the order $\left|\omega_{0}\right|\approx1.72$ that is superimposed
on the function. In the corresponding plots of $\tilde{C}_{n}^{\left(R\right)}=Re\left(\tilde{C}_{n}\left(\tau\right)\right)$,
presented in Fig.\ref{fig:Power Spectrum + A_C_function}.d and Fig.\ref{fig:Power Spectrum + A_C_function}.e,
one does not find this oscillation. Behavior of the imaginary part
of the auto-correlation function $\widetilde{C}_{n}^{\left(I\right)}=Im\left(\widetilde{C}_{n}\left(\tau\right)\right)$
is similar (see Fig.\ref{fig:Power Spectrum + A_C_function}.f). All
results presented in Fig.\ref{fig:Power Spectrum + A_C_function}
are for $n=0.$ Similar results were found also for $n=3$ and $n=15$.
We see that the auto-correlation function decays by 2 orders of magnitude
on the scale of $\triangle\tau\approx140$ (of the order of $2\pi/\triangle\omega\sim65$).
Therefore the correlation of $\tilde{F}_{n}\left(t\right)$ behaves
as the one of noise with short time correlations. For realizations
where the growth of the second moment $M_{2}\sim t^{1/3}$ was not
found, the power spectrum was found to be substantially narrower by
2 orders of magnitude. The calculations were repeated for $\beta=2$
where similar results were found, and for $\beta=0.5$. For the latter
case the number of realizations where it was found that the second
moment grows like $t^{1/3}$ is substaintially smaller than for $\beta=1$
or $\beta=2$. In all cases where the width of the power spectrum
was small the typical growth of the secomd moment $M_{2}\sim t^{1/3}$
was not found and vice versa. \emph{This demonstrates the strong relation
between the effective noise behavior and the diffusive growth of the
second moment. It also demonstrates the different behavior of various
realizations of the randomness.} 

\begin{figure}
\includegraphics[scale=0.45]{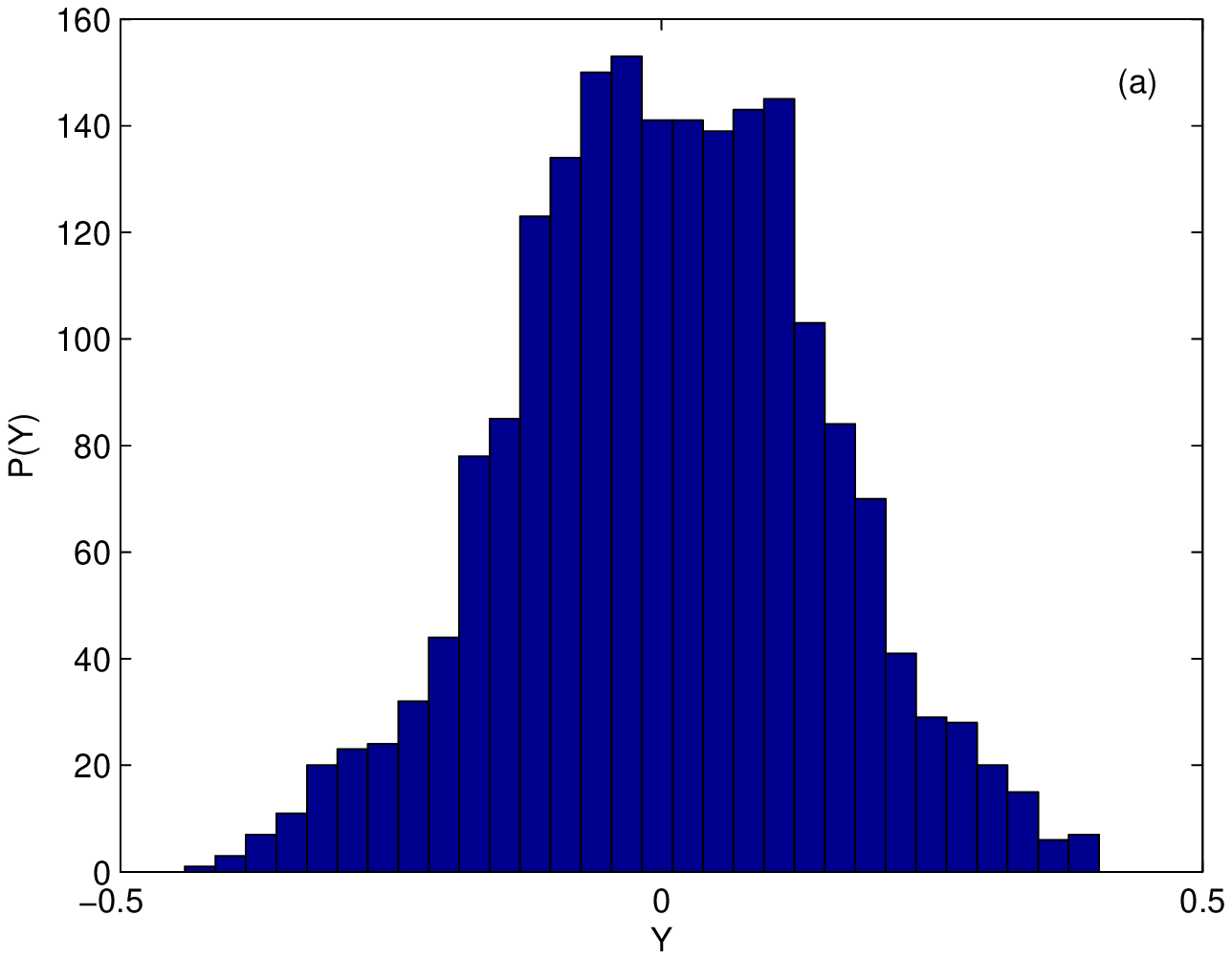}\includegraphics[scale=0.45]{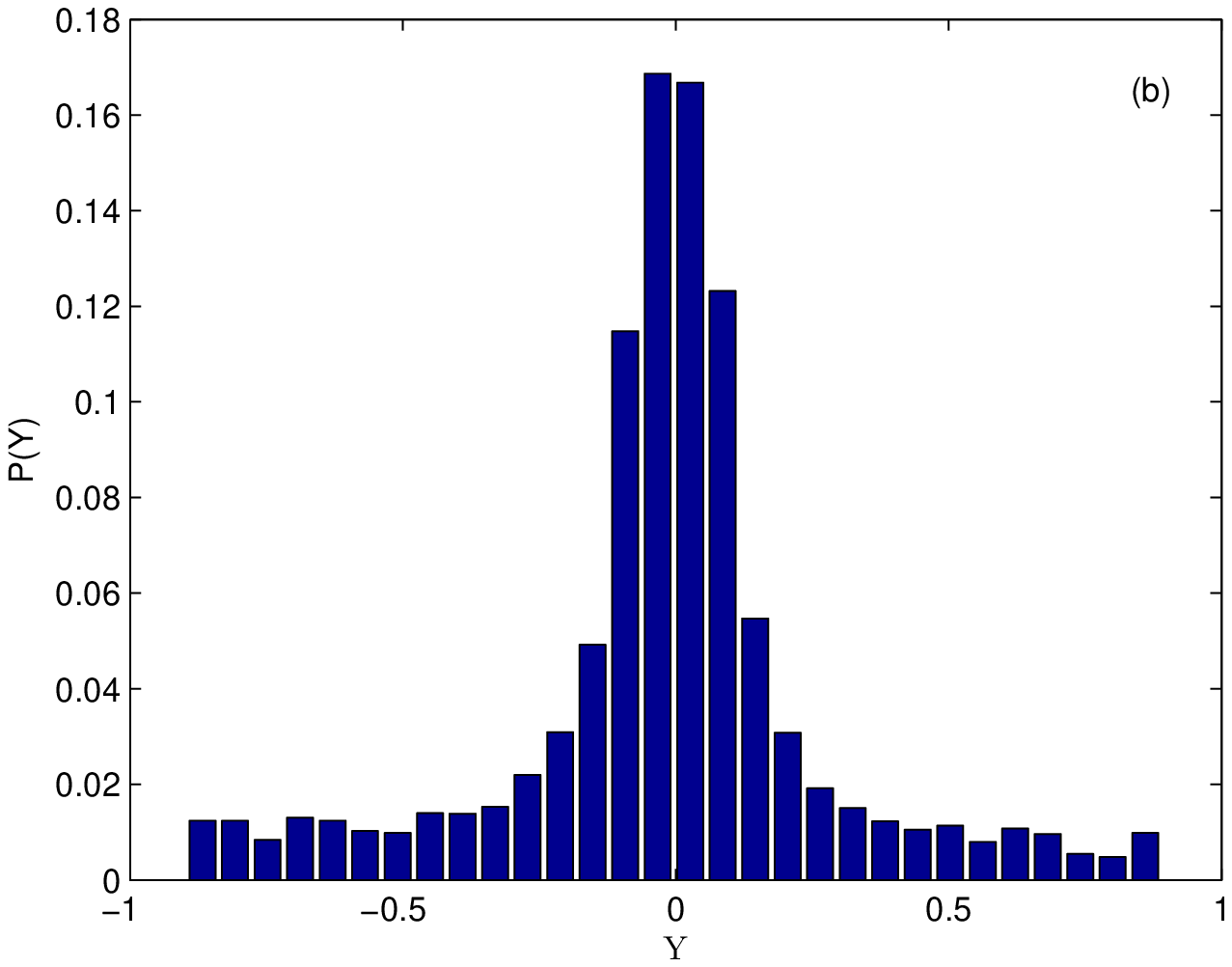}\caption{\label{fig:Distribution} The distribution of $Y=\tilde{F}_{n}^{\left(R\right)}\left(k\cdot t_{a}\right)$
where $k=(1,2,..K)$ , $K=500$, $t_{a}=200$ , $t=10^{5}$ and the
bin size $0.0596$ . (a) For the same realization used in Fig. 1 .
(b) The distribution of values found for all $N_{R}=50$ realizations.}
\end{figure}

We turn now to test the distribution of $\tilde{F}_{n}\left(t\right)$.
For this purpose we sample $\tilde{F}_{n}\left(t\right)$ for a sequence
of points separated by $t_{a}>\triangle\tau$ , that is for points
where the values of $\tilde{F}_{n}\left(t\right)$ are uncorrelated,
and compute the distribution of $\tilde{F}_{n}\left(k\cdot t_{a}\right)$
for $k=(1,2,..K)$. The results are presented in Fig. \ref{fig:Distribution}
for $t=10^{5}$ , $t_{a}=200$, $K=500.$

\subsection{\label{sub:3.2}Estimate of scaling of the matrix elements $V_{n}^{m_{1},m_{2},m_{3}}$
with $\xi$}

\begin{figure}
\includegraphics[scale=0.5]{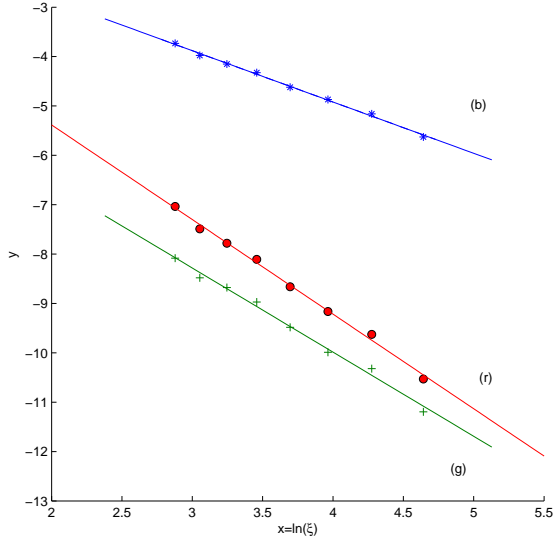}\caption{\label{fig:A-log-log-plot_V}A log-log plot of (b) $y=ln\left\langle V_{0}^{0,\frac{\xi}{3},\frac{\xi}{3}}\right\rangle $,
(r) $y=ln\left\langle \left(V_{0}^{0,\frac{\xi}{3},\frac{\xi}{3}}\right)^{2}\right\rangle $
and (g) $y=ln\left(\left\langle \left(V_{0}^{0,\frac{\xi}{3},\frac{\xi}{3}}\right)^{2}\right\rangle -\left\langle V_{0}^{0,\frac{\xi}{3},\frac{\xi}{3}}\right\rangle ^{2}\right)$
as a function of $x=ln\mbox{\ensuremath{\left(\xi\right)} }$, for
the parameters $N=512$ , $N_{R}=5000$. The localization length varies
in the interval $11<\xi<103.$ The least square fit leads to $\eta_{1}=1.039$
, $\eta_{2}=0.958$ and $\eta_{3}=0.853$ respectively .}
\end{figure}

The overlap sum $V_{n}^{m_{1},m_{2},m_{3}}$ is a random function.
In this subsection the scaling of its typical values with the maximal
localization length \cite{DerridaB.1984} 
\begin{equation}
\xi\approx\frac{96}{W^{2}}\label{eq:approx xi}
\end{equation}
 is evaluated. This relation holds in the limit of weak disorder.
In the numerical calculations presented in this paper we vary $W$
as the control parameter and the localization length is calculated
from \eqref{eq:approx xi}. The estimate \eqref{eq:approx xi} is
a reasonable approximation for $W<5.5$ or $\xi>3.15$ as was checked
explicitly (and used) in this subsection. We note that the $V_{n}^{m_{1},m_{2},m_{3}}$
take values of substantial magnitude when all the centers of localization
of the states $u_{n},u_{m_{1}},u_{m_{2}},u_{m_{3}}$ are within a
distance $\xi.$ Only such overlap sums are considered. The average
of the overlap sums over realizations vanishes unless $(n,m_{1},m_{2},m_{3})$
consists of two pairs of identical values , $n=m_{1}$ and $m_{2}=m_{3}$
and all permutations. We calculated $\left\langle \left|V_{n}^{m_{1},m_{2},m_{3}}\right|^{2}\right\rangle $
and $\left\langle V_{n}^{m_{1},m_{2},m_{3}}\right\rangle $ (where
$\left\langle \cdot\right\rangle $ denotes average over $N_{R}=5000$
realizations) while $x_{n},x_{m_{1}},x_{m_{2}},x_{m_{3}}$ are fixed
fractions of $\xi$, while $\xi$ (and $W$) are varied. Assuming
$\left\langle V_{n}^{m_{1},m_{2},m_{3}}\right\rangle \thicksim\xi^{-\eta_{1}}$
and $\left\langle \left|V_{n}^{m_{1},m_{2},m_{3}}\right|^{2}\right\rangle \thicksim\xi^{-2\eta_{2}}$
while the variance $\left\langle \left(V_{n}^{m_{1},m_{2},m_{3}}\right)^{2}\right\rangle -\left\langle V_{n}^{m_{1},m_{2},m_{3}}\right\rangle ^{2}$
scales as $\xi^{-2\eta_{3}}$, we estimate these exponents from Figures
like Fig. \ref{fig:A-log-log-plot_V} . We conclude that $\eta_{1}\thickapprox\eta_{2}\thickapprox\eta_{3}\thickapprox1$
. Therefore the typical magnitude of the random variable $V_{n}^{m_{1},m_{2},m_{3}}$
scales as \eqref{Vt} with $\eta=1$. Although this result is expected
from the scaling theory of localization, it is not obvious appriory.
In particular it is not clear what is the effect of cancellations
of various terms resulting of opposite signs.

For $\xi\ll11$ we could not obtain smooth curves of $V_{n}^{m_{1},m_{2},m_{3}}$.
The reason is that the centers of localization $x_{m_{i}}$ are equal
to the integer part of $\nicefrac{\xi}{a}$ where $a$ is fixed and
$\xi$ varies. For small $\xi$ the jumps in $V_{n}^{m_{1},m_{2},m_{3}}$
are significant, since $\xi$ does not cover many integers. The results
obtained indicate that scaling of the overlap sums as $\xi^{-1}$
holds also for values $\xi<11$. In summary for a crude evaluation
one can assume (\ref{Vt}) holds with $\eta=1$.

\subsection{\label{sub:3.3}The scaling of the second moment $M_{2}$ with $\xi$
( and $\beta$)}

\begin{figure}
\includegraphics[scale=0.5]{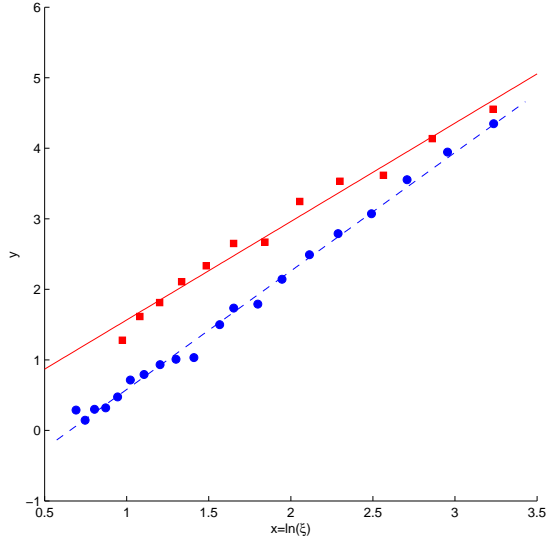}

\caption{\label{fig:-from-beta=00003D1}The dependence of $A$ defined by \eqref{eq:m2}
and \eqref{eq:A4} for $\beta=1$ (blue circles) and for $\beta=3$
(red squares) on $\xi$. We denote $y=ln\left(A\right)$ and $x=ln\mbox{\ensuremath{\left(\xi\right)} }$.
From the least square fit we find $\nu=1.684$ for $\beta=1$ (blue)
and $\nu=1.395$ for $\beta=3$ (red).}
\end{figure}

In this subsection we will estimate the exponent $\alpha$ defined
in \eqref{P}. For this purpose we write \eqref{eq:m2-1} in the form
\begin{equation}
M_{2}=At^{\frac{1}{3}}\label{eq:m2}
\end{equation}
with 
\begin{equation}
A=A_{4}\xi^{\nu}\label{eq:A4}
\end{equation}
where $\nu=\frac{2}{3}\left(\alpha-\eta+1\right)$ (see \eqref{eq:18})
while $A_{4}$ is a constant independent of $\xi$. We used the split
step method to obtain $\psi\left(x,t\right)$ for different realizations
($N_{R}=30)$ and computed $\psi$ until $t=10^{6}$ . Only realizations
which satisfied $M_{2}\sim t^{\frac{1}{3}}$ at some stage of the
calculation were taken into account. This was the case for nearly
all the $N_{R}$ realizations for $\xi>7$ and $\beta<4$. In the
other regimes it was not satisfied for a significant number of realizations.
Fixing $\beta$ we estimate $\nu$ from plots like Fig. \ref{fig:-from-beta=00003D1}.
For $1<\beta<3.5$ using the fact that $\eta\thickapprox1$ we find
that for $1.235<\nu<1.71$ for various values of $\beta.$ The exponent
$\alpha$ of \eqref{P} takes the values $1.85<\alpha<2.56$ . We
note the strong uncertainty of $\nu$ and $\alpha$. These results
indicate that $A$$\sim\xi^{\nu}$. It is an estimate of the order
of magnitude but not a verification of this power law. 

\section{\label{sec:Possobolity-for-the-breakdown}Possibility for the breakdown
of the effective noise theory}

For the effective noise theory it is essential that $F_{n}\left(t\right)$
can be considered random. For this the number of terms in the sum
(\ref{Cs}) that resonate with $n$ should be large,namely $\mathcal{P}$
should not be too small. The density $\rho$ and therefore $\mathcal{P}$
decrease with time. If $\mathcal{P}$ is very small there may be a
situation that as a result of fluctuations, the sum (\ref{Cs}) is
dominated just by one term and therefore it is effectively quasi periodic.
If spreading is a result of the randomness of $F_{n}$, it will stop
then. Let us first estimate the time scale required to spread so that
$\mathcal{P}\approx1$. For this purpose let us write (\ref{P}) in
the form
\begin{equation}
\mathcal{P}\thickapprox\overline{A}\xi^{\alpha}\rho\label{eq:ro}
\end{equation}
where $\overline{A}=A_{0}\beta$. Since $\rho$ decreases with time
$t$ there is a time scale when $\mathcal{P}$ will become very small.
Assuming the constants are of the order of unity, using \eqref{eq:18}
and \eqref{eq:ro_D_t} the time $t^{*}$ when $\mathcal{P}\approx1$
satisfies

\begin{equation}
\xi^{2\alpha}\cdot\frac{1}{\left[\xi^{2\left(\alpha-\eta+1\right)}t^{*}\right]^{\frac{1}{3}}}\approx1
\end{equation}
or 
\begin{equation}
\xi^{\left(\frac{4}{3}\alpha+\frac{2}{3}\left(\eta-1\right)\right)}\approx t^{*^{\frac{1}{3}}}\label{eq:t*}
\end{equation}
 resulting in 
\begin{equation}
t^{*}\approx\xi^{\left(4\alpha+2\left(\eta-1\right)\right)}
\end{equation}
 for $1.85<\alpha<2.56$ and $\eta=1$ 
\begin{equation}
t^{*}\approx\xi^{\delta}
\end{equation}
 where $7.4<\delta<10.24$\\
The time required for, $\mathcal{P}\ll1$, when the effective noise
theory may fail is even larger.

\section{Summary and conclusions}

The effective noise theory was introduced in \cite{Shepelyansky1993}
and was further developed in \cite{Flach2009,Pikovsky2008,Skokos2009}.
It was found to be consistent with the numerical results in some regimes.
In Section \ref{sec:The-effective-noise} our interpretation of this
theory was presented. In section \ref{sec:Numerical-tests-for} the
details of this theory were tested numerically. In particular the
distribution of the effective driving $F_{n}$ defined in \eqref{Cs}
was studied . The correlation function was calculated as well and
was found to be characterized by a wide power spectrum and rapid decay
with time. These were found only for realizations where subdiffusion
with the second moment growing as $t^{1/3}$ is found, indicating
the relation between this spreading and the approximation of $F_{n}$
as effective noise. These results are purely numerical and support
the effective noise theory. An obvious challenge is to obtain these
results analytically. We determined that the behavior $A\approx\xi^{\nu}$
(see \eqref{eq:A4}), with $1.235<\nu<1.71$ is a reasonable approximation.
From this we conclude that the dependence of $\mathcal{P}$ on $\xi$
\eqref{P} is controlled by the exponent $1.85<\alpha<2.56$. Although
$\xi$ varied over one decade and the evaluation of the exponent is
crude we believe it may give the correct order of magnitude.

We turn to speculate how the effective noise theory may break down
for a long time scale. Assuming the effective noise theory holds for
long time, $\mathcal{P}$ of \eqref{P} becomes extremely small, consequently
the number of terms in the sum \eqref{Cs} that contribute significantly
may become of order unity and $F_{n}$ may turn to be quasi periodic
rather that random. Therefore there is a time scale $t^{*}$ given
by the estimate \eqref{eq:t*} so that for $t>t^{*}$ the effective
noise theory is invalid. For such long time a sequence of peaks may
replace the continuous region of the power spectrum in Fig 1.a. If
localization is destroyed by the effective noise $F_{n}$, it is reasonable
to expect localization or spreading slower than subdiffusion (say
logarithnic in time) on time scale $t^{*}$ and larger. Existence
of such a time scale is consistent with \cite{Wang2008,Wang2008a,Fishman2009a,Pikovsky2011,Johansson2010,Fishman2008a}.
The scaling arguments used here should improve when the localization
length $\xi$ becomes large but then $t^{*}$ becomes extremely large
and it is impossible to explore numerically the scenario for the breakdown
of the effective noise theory outlined in Sec. \ref{sec:Possobolity-for-the-breakdown}.
Such a scenario may enable to reconcile the numerical results where
subdiffusion is found \cite{Fishman2011,Pikovsky2008,Kopidakis2008,Molina1998,Flach2009,Skokos2009}
with the analytical results predicting asymptotic spreading that is
at most logarithmic \cite{Fishman2011,Wang2008,Wang2008a,Fishman2009a}.
These points should be subject of future research.

\paragraph{Acknowledgment}

We would like to thank Y. Krivolapov for detailed discussions, extremely
valuable technical detailed help and for extremely critical reading
of the menuscript . We would like to thank J. Bodyfelt, S. Flach,
D. Krimer, A. Pikovsky and A. Soffer for useful discussions. We thank
a referee of Physical Review for suggesting the argument in the end
of Sec.2 leading to $\gamma=1$. This work was partly supported by
the Israel Science Foundation (ISF), by the US-Israel Binational Science
Foundation (BSF), by the Minerva Center of Nonlinear Physics of Complex
Systems, by the New York Metropolitan Research Fund and by the Shlomo
Kaplansky academic chair.

\section*{Appendix: some details of the numerical calculations}

We used the split step method to obtain the time evolution starting
from the initial wavefunction. The lattice size $N$ used is $512$
or $1024$. The reason we used the relativity large lattice is because
we wanted to avoid boundary effects, namely we required the wavefnction
amplitude to be smaller than $10^{-12}$ on the boundary. The time
step used in the split step method is $dt=0.1$. We used this time
step because it is small enough relative to the time scales in the
system at hand and large enough in order to complete the numerical
calculation in reasonable time. It is the smallest time step used
in \cite{Flach2009,Skokos2009}. The initial condition used is a single
site excitation in the middle of the lattice denoted by $x_{n}=0$
namely, $\psi\left(x,t=0\right)=\delta_{x,0}$.

\bibliographystyle{unsrt}
\bibliography{NLSE}

\begin{thebibliography}{10}

\bibitem{Sulem1999}
C.~Sulem and P.~L. Sulem.
\newblock {\em The nonlinear Schr\"{o}dinger equation self-focusing and wave
  collapse}.
\newblock Springer, 1999.

\bibitem{Agrawal2007}
G.~P. Agrawal.
\newblock {\em Nonlinear fiber optics}, volume 4th.
\newblock Academic Press, Burlington, MA ; London, 2007.

\bibitem{Dalfovo1999}
F.~Dalfovo, S.~Giorgini, L.~P. Pitaevskii, and S.~Stringari.
\newblock Theory of {B}ose-{E}instein condensation in trapped gases.
\newblock {\em Rev. Mod. Phys.}, 71(3):463--512, 1999.

\bibitem{Pitaevskii2003}
L.~P. Pitaevskii and S.~Stringari.
\newblock {\em {B}ose-{E}instein condensation}.
\newblock Clarendon Press, Oxford ; New York, 2003.

\bibitem{Leggett2001}
A.~J. Leggett.
\newblock {B}ose-{E}instein condensation in the alkali gases: {S}ome
  fundamental concepts.
\newblock {\em Rev. Mod. Phys.}, 73(2):307--356, 2001.

\bibitem{Pitaevskii1961}
L.P. Pitaevskii.
\newblock Vortex lines in an imperfect {B}ose gas.
\newblock {\em JETP}, 13(2):451--454, 1961.

\bibitem{Gross1961}
E.P. Gross.
\newblock Structure of a quantized vortex in boson systems.
\newblock {\em Nuovo Cimento}, 20(3):454--477, 1961.

\bibitem{Ishii1973}
K.~Ishii.
\newblock Localization of eigenstates and transport phenomena in
  one-dimensional disordered system.
\newblock {\em Suppl. Prog, Theor. Phys.}, 53(53):77--138, 1973.

\bibitem{Lee1985}
P.~A. Lee and T.~V. Ramakrishnan.
\newblock Disordered electronic systems.
\newblock {\em Rev. Mod. Phys.}, 57(2):287--337, 1985.

\bibitem{Lifshits1988}
I.~M. Lifshits, L.~A. Pastur, and S.~A. Gredeskul.
\newblock {\em Introduction to the theory of disordered systems}.
\newblock Wiley, New York, 1988.

\bibitem{Anderson1958}
P.~W. Anderson.
\newblock {A}bsence of diffusion in certain random lattices.
\newblock {\em Phys. Rev.}, 109(5):1492, 1958.

\bibitem{Schwartz2007}
T.~Schwartz, G.~Bartal, S.~Fishman, and M.~Segev.
\newblock Transport and {A}nderson localization in disordered two-dimensional
  photonic lattices.
\newblock {\em Nature}, 446(7131):52--55, 2007.

\bibitem{Lahini2008}
Y.~Lahini, A.~Avidan, F.~Pozzi, M.~Sorel, R.~Morandotti, D.~Christodoulides,
  and Y.~Silberberg.
\newblock Anderson localization and nonlinearity in one-dimensional disordered
  photonic lattices.
\newblock {\em Phys. Rev. Lett.}, 100(1):013906, Jan 2008.

\bibitem{Gimperlein2005}
H.~Gimperlein, S.~Wessel, J.~Schmiedmayer, and L.~Santos.
\newblock Ultracold atoms in optical lattices with random on-site interactions.
\newblock {\em Phys. Rev. Lett.}, 95(17):170401, 2005.

\bibitem{Lye2005}
J.~E. Lye, L.~Fallani, M.~Modugno, D.~S. Wiersma, C.~Fort, and M.~Inguscio.
\newblock {B}ose-{E}instein condensate in a random potential.
\newblock {\em Phys. Rev. Lett.}, 95(7):070401, 2005.

\bibitem{Clement2005}
D.~Clement, A.~F. Varon, M.~Hugbart, J.~A. Retter, P.~Bouyer,
  L.~Sanchez-Palencia, D.~M. Gangardt, G.~V. Shlyapnikov, and A.~Aspect.
\newblock Suppression of transport of an interacting elongated
  {B}ose-{E}instein condensate in a random potential.
\newblock {\em Phys. Rev. Lett.}, 95(17):170409, 2005.

\bibitem{Clement2006}
D.~Clement, A.~F. Varon, J.~A. Retter, L.~Sanchez-Palencia, A.~Aspect, and
  P.~Bouyer.
\newblock {E}xperimental study of the transport of coherent interacting
  matter-waves in a 1{D} random potential induced by laser speckle.
\newblock {\em New J. Phys.}, 8:165, 2006.

\bibitem{Sanchez-Palencia2007}
L.~Sanchez-Palencia, D.~Clement, P.~Lugan, P.~Bouyer, G.~V. Shlyapnikov, and
  A.~Aspect.
\newblock Anderson localization of expanding {B}ose-{E}instein condensates in
  random potentials.
\newblock {\em Phys. Rev. Lett.}, 98(21):210401, May 2007.

\bibitem{Billy2008}
J.~Billy, V.~Josse, Z.~C. Zuo, A.~Bernard, B.~Hambrecht, P.~Lugan, D.~Clement,
  L.~Sanchez-Palencia, P.~Bouyer, and A.~Aspect.
\newblock Direct observation of {A}nderson localization of matter waves in a
  controlled disorder.
\newblock {\em Nature}, 453(7197):891--894, June 2008.

\bibitem{Fort2005}
C.~Fort, L.~Fallani, V.~Guarrera, J.~E. Lye, M.~Modugno, D.~S. Wiersma, and
  M.~Inguscio.
\newblock {E}ffect of optical disorder and single defects on the expansion of a
  {B}ose-{E}instein condensate in a one-dimensional waveguide.
\newblock {\em Phys. Rev. Lett.}, 95(17):170410, 2005.

\bibitem{Akkermans2008}
E.~Akkermans, S.~Ghosh, and Z.~H. Musslimani.
\newblock Numerical study of one-dimensional and interacting {B}ose-{E}instein
  condensates in a random potential.
\newblock {\em J. Phys. B}, 41(4):045302, 2008.

\bibitem{Paul2007}
T.~Paul, P.~Schlagheck, P.~Leboeuf, and N.~Pavloff.
\newblock Superfluidity versus {A}nderson localization in a dilute {B}ose gas.
\newblock {\em Phys. Rev. Lett.}, 98(21):210602, 2007.

\bibitem{Beilin2010}
L.~Beilin, E.~Gurevich, and B.~Shapiro.
\newblock Diffusion of cold-atomic gases in the presence of an optical speckle
  potential.
\newblock {\em Phys. Rev. A}, 81(3):033612, Mar 2010.

\bibitem{Bishop1995}
A.~R. Bishop.
\newblock {\em Fluctuation phenomena : disorder and nonlinearity}.
\newblock World Scientific, Singapore ; River Edge, NJ, 1995.

\bibitem{Rasmussen1999}
K.~O. Rasmussen, D.~Cai, A.~R. Bishop, and N.~Gronbech-Jensen.
\newblock Localization in a nonlinear disordered system.
\newblock {\em Europhys. Lett.}, 47(4):421--427, 1999.

\bibitem{Kopidakis1999}
G.~Kopidakis and S.~Aubry.
\newblock Intraband discrete breathers in disordered nonlinear systems. {I.
  D}elocalization.
\newblock {\em Physica D}, 130(3-4):155--186, 1999.

\bibitem{Kopidakis2000}
G.~Kopidakis and S.~Aubry.
\newblock Discrete breathers and delocalization in nonlinear disordered
  systems.
\newblock {\em Phys. Rev. Lett.}, 84(15):3236--3239, 2000.

\bibitem{Fishman2011}
S.~Fishman, Y.~Krivolapov, and A.~Soffer.
\newblock The nonlinear schr\"{o}dinger equation with random potential :
  Results and puzzles . arxiv:1108.2956 to be published in nonlinearity.
\newblock 2011.

\bibitem{Wang2008}
W.-M. Wang and Z.~Zhang.
\newblock Long time {A}nderson localization for nonlinear random
  {S}chr\"{o}dinger equation.
\newblock {\em J. Stat. Phys.}, 134:953, 2009.

\bibitem{Wang2008a}
W.-M. Wang.
\newblock Logarithmic bounds on {S}obolev norms for time dependent linear
  {S}chr\"{o}inger equations.
\newblock {\em Comm. Part. Diff. Eq.}, 33(12):2164--2179, 2008.

\bibitem{Fishman2009a}
S.~Fishman, Y.~Krivolapov, and A.~Soffer.
\newblock Perturbation theory for the nonlinear {S}chr\"{o}dinger equation with
  a random potential.
\newblock {\em Nonlinearity}, 22:2861--2887, 2009.

\bibitem{Krivolapov2010}
Y.~Krivolapov, S.~Fishman, and A.~Soffer.
\newblock A numerical and symbolical approximation of the nonlinear {A}nderson
  model.
\newblock {\em New J. Phy.}, 12(6):063035, 2010.

\bibitem{Pikovsky2011}
A.~Pikovsky and S.~Fishman.
\newblock Scaling properties of weak chaos in nonlinear disordered lattices.
\newblock {\em Phys. Rev. E}, 83(2):025201, Feb 2011.

\bibitem{Ivanchenko2011}
M.~V Ivanchenko, T.~V Laptyeva, and S.~Flach.
\newblock Anderson localization or nonlinear waves? a matter of probability.
\newblock arXiv:1108.0899v1, 2011.

\bibitem{Benettin1988}
G.~Benettin, J.~Fr\"ohlich, and A.~Giorgilli.
\newblock A {N}ekhoroshev-type theorem for {H}amiltonian-systems with
  infinitely many degrees of freedom.
\newblock {\em Commun. Math. Phys.}, 119(1):95--108, 1988.

\bibitem{Johansson2010}
M.~Johansson, G.~Kopidakis, and S.~Aubry.
\newblock {KAM} tori in 1{D} random discrete nonlinear {S}chr\"{o}inger model?
\newblock {\em EPL (Europhysics Letters)}, 91(5):50001, 2010.

\bibitem{Basko2011}
D.~M. Basko.
\newblock Weak chaos in the disordered nonlinear {S}chr\"{o}dinger chain:
  Destruction of {A}nderson localization by {A}rnold diffusion.
\newblock {\em Annal. Phys.}, 326(7):1577--1655, 2011.

\bibitem{Frohlich1986}
J.~Fr\"{o}hlich, T.~Spencer, and C.~E. Wayne.
\newblock Localization in disordered, nonlinear dynamic-systems.
\newblock {\em J. Stat. Phys.}, 42(3-4):247--274, 1986.

\bibitem{Shepelyansky1993}
D.~L. Shepelyansky.
\newblock Delocalization of quantum chaos by weak nonlinearity.
\newblock {\em Phys. Rev. Lett.}, 70(12):1787--1790, 1993.

\bibitem{Pikovsky2008}
A.~S. Pikovsky and D.~L. Shepelyansky.
\newblock Destruction of {A}nderson localization by a weak nonlinearity.
\newblock {\em Phys. Rev. Lett.}, 100(9):094101, 2008.

\bibitem{Kopidakis2008}
G.~Kopidakis, S.~Komineas, S.~Flach, and S.~Aubry.
\newblock Absence of wave packet diffusion in disordered nonlinear systems.
\newblock {\em Phys. Rev. Lett.}, 100(8):084103, 2008.

\bibitem{Molina1998}
M.~I. Molina.
\newblock Transport of localized and extended excitations in a nonlinear
  {A}nderson model.
\newblock {\em Phys. Rev. B}, 58(19):12547--12550, 1998.

\bibitem{Flach2009}
S.~Flach, D.~Krimer, and Ch. Skokos.
\newblock Universal spreading of wavepackets in disordered nonlinear systems.
\newblock {\em Phys. Rev. Lett.}, 102:024101, 2009.

\bibitem{Skokos2009}
C.~Skokos, D.O. Krimer, Komineas, and S.~S.~Flach.
\newblock Delocalization of wave packets in disordered nonlinear chains.
\newblock {\em Phys. Rev. E}, 79:056211, 2009.

\bibitem{Fishman2008a}
S.~Fishman, Y.~Krivolapov, and A.~Soffer.
\newblock On the problem of dynamical localization in the nonlinear
  {S}chr\"{o}dinger equation with a random potential.
\newblock {\em J. Stat. Phys.}, 131(5):843--865, 2008.

\bibitem{Basko}
D.~Basko.
\newblock Weak chaos in the disordered nonlinear {S}chr\"{o}dinger chain:
  Destruction of {A}nderson localization by {A}rnold diffusion.
\newblock {\em Annal. Phys.}, 326:1577--1655, 2011.

\bibitem{DerridaB.1984}
B.~Derrida and E.~Gardner.
\newblock Lyapounov exponent of the one dimensional anderson model : weak
  disorder expansions.
\newblock {\em J. Phys. France}, 45(8):1283--1295, 1984.

\bibitem{MacKinnon1981}
A.~MacKinnon and B.~Kramer.
\newblock One-parameter scaling of localization length and conductance in
  disordered systems.
\newblock {\em Phys. Rev. Lett.}, 47(21):1546--1549, Nov 1981.

\end{thebibliography}

\end{document}